\documentclass[proceedings]{JHEP} % 10pt is ignored!
\usepackage{epsfig}                     % please use epsfig.
\newbox\mybox
\newcommand{\sss}{\scriptscriptstyle}
\newcommand\fverb{\setbox\mybox=\hbox\bgroup\verb}
\newcommand\fverbdo{\egroup\medskip\noindent\fbox{\unhbox\mybox}\ }
\newcommand\fverbit{\egroup\item[\fbox{\unhbox\mybox}]}

\font\beeg=cmr17 scaled 1600            % Stylish initials
\newcommand\init[1]{\setbox\mybox=\hbox{{\beeg #1}~}%
                   \noindent\global\hangindent=\wd\mybox\global\hangafter-2%
                   \sc\smash{\llap {\lower 13.2pt \box\mybox}}}
\title{Theory of Nonleptonic $B$ Decays}

\author{L. Silvestrini\\
        Physik-Department, Technische Universit{\"a}t M{\"u}nchen,
        D-85747 Garching, Germany\\
        E-mail: \email{Luca.Silvestrini@Physik.TU-Muenchen.DE}}

\conference{Heavy Flavours 8, Southampton, UK, 1999}

\abstract{I briefly review some recent progress in the theory of
  nonleptonic $B$ decays. After introducing the operator product
  expansion and the relevant effective Hamiltonian, I discuss the
  domain of validity and the theoretical justification of the
  factorization approximation and of its generalizations. Furthermore,
  I review some general parameterizations of $B$ decay amplitudes: the
  ``diagrammatic'' approach and the formalism based on Wick
  contractions in the matrix elements of four-fermion operators.}

\begin{document} 

\maketitle %%%%%%%%%% THIS IS IGNORED %%%%%%%%%%%

\section{Introduction}
\label{sec:intro}

The theoretical description of nonleptonic $B$ decays is an extremely
difficult problem, due to the nonperturbative dynamics connected with
hadronic final states. However, a good understanding of these
transitions, or at least a reliable estimate of the theoretical
uncertainties, is necessary to give a correct interpretation to the
new data that are being collected at dedicated experiments. In
particular, a meaningful test of the standard model and of its
extensions on the most promising ground of CP violation in $B$ decays 
requires a good theoretical description of these processes.

The starting point of the analysis is the operator product expansion,
that allows us to write the transition amplitude from a $B$ meson into
a final state $F$ in terms of perturbative Wilson coefficients and
nonperturbative hadronic matrix elements of local operators:
\begin{eqnarray}
  \label{eq:ope}
  && \mathcal{A}(B \to F) = \langle F \vert \mathcal{H}_\mathrm{eff}
  \vert B \rangle = \nonumber \\
  && \frac{G_F}{\sqrt{2}} \sum_i V^{\mathrm{CKM}}_i
  C_i(\mu) \langle F \vert Q_i (\mu) \vert B \rangle,
\end{eqnarray}
where $\mu$ is the renormalization scale. The Wilson coefficients
$C_i(\mu)$ and the matrix elements $\langle Q_i(\mu)\rangle$ are
individually renormalization scale and scheme dependent, but they
combine in eq.~(\ref{eq:ope}) to give a scale and scheme independent
result for the amplitude (up to a residual scale dependence of higher
order in the perturbative expansion). The $C_i(\mu)$ are universal,
while all the dependence on the external states is carried by the
matrix elements. 

Wilson coefficients can be reliably computed in perturbation
theory, and the full next-to-leading order results for
$\mathcal{H}_\mathrm{eff}^{\Delta B=1}$ have been computed a few years
ago by the Munich and Rome groups \cite{NLOQCD}.

On the other hand, the computation of matrix elements requires the use
of some nonperturbative technique. Unfortunately, a model independent
computation of $\langle Q_i(\mu)\rangle$ from first principles on the
lattice is not possible due to the Maiani-Testa no-go theorem
\cite{mtngt}. A method to extract $\langle Q_i(\mu)\rangle$ from
lattice QCD in a model-dependent way has been proposed \cite{lattFSI},
but its feasibility still has to be verified. Light-cone
QCD sum rules might be used to estimate the matrix elements, however
Final State Interaction (FSI) phases cannot be computed in this way
\cite{braunhf8}. It is therefore fair to say that at the moment no
method is available to compute nonleptonic $B$ decays from first
principles. One is then left with two possibilities. 

The first is to use some approximation to simplify the dynamics to
obtain an estimate of the matrix elements. Factorization is the
simplest example of such approximations and it has been extensively
used to study nonleptonic $B$ decays. The advantage of this kind of
approach is the possibility of describing a large class of decay
channels with few parameters.  However, as we shall see in the
following, factorization holds only in some channels and up to
power-suppressed corrections.  These corrections might, for example,
play an important role in the extraction of CKM parameters from
CP-violating $B$ decays. To estimate the possible effects of these
power-suppressed terms, and to correctly describe those channels in
which factorization cannot be justified, it is useful to develop
another formalism. One can construct a general parameterization of
weak decay amplitudes, including all possible hadronic dynamics. This
can be supplemented with some symmetry argument to gain some
predictive power, and when enough data are available, one can try to
extract the hadronic parameters from the experiment.

In the following sections I will briefly summarize
some recent progress that has been made in these two directions,
discussing the advantages and disadvantages of these complementary
approaches.

\section{Factorization}
\label{sec:fact}

The factorization approximation consists in writing the matrix element
of a four-quark operator between the $B$ meson and a two-body hadronic
final state as the product of two matrix elements of quark bilinears
\cite{fact}--\cite{NEUBERT}. For example, let us consider the decay
$\bar B^0 \to D^+ \pi^-$, which is mediated by the following
operators:
\begin{displaymath}
 Q_1=(\bar d u)_{V-A} (\bar c
b)_{V-A},\; Q_2=(\bar d b)_{V-A} (\bar c
u)_{V-A}.
\end{displaymath}
In factorization, the matrix elements of $Q_1$ and $Q_2$ are written as 
\begin{eqnarray}
  &&\langle \pi^- D^+ \vert Q_1 \vert
  \bar B^0 \rangle_\mathrm{F} = \langle \pi^- \vert (\bar d u)_{A}
  \vert 0 \rangle 
  \langle D^+ \vert   (\bar c b)_V \vert \bar B^0 \rangle, \nonumber \\
  &&\langle \pi^- D^+ \vert Q_2 \vert
  \bar B^0 \rangle_\mathrm{F} = \frac{1}{N} \langle \pi^- D^+ \vert Q_1 \vert
  \bar B^0 \rangle_\mathrm{F} +\nonumber \\ 
   && \frac{1}{2} \langle \pi^- D^+ \vert
  (\bar d t^a u)_{V-A} (\bar c t^a b)_{V-A} \vert \bar B^0 \rangle
  \nonumber \\
  && \qquad \to \frac{1}{N} \langle \pi^- D^+ \vert Q_1 \vert
  \bar B^0 \rangle_\mathrm{F},
  \label{eq:fact1}
\end{eqnarray}
where $t^a$ is a colour matrix, $N$ the number of colours and the
octet-octet term in $\langle Q_2 \rangle$ has been put to zero. From
eq.~(\ref{eq:fact1}) it is evident that in factorization final state
interactions and gluon exchanges between the two currents are fully
neglected. From eq.~(\ref{eq:fact1}) one gets
\begin{eqnarray}
  \label{eq:bdp}
  &&\langle \pi^- D^+ \vert \mathcal{H}_\mathrm{eff} \vert
  \bar B^0 \rangle_\mathrm{F}\propto  \\ 
  &&\left(C_1(\mu) + \frac{1}{N}
  C_2(\mu)\right) \times 
  \langle \pi^- D^+ \vert Q_1 \vert
  \bar B^0 \rangle_\mathrm{F}. \nonumber
\end{eqnarray}
The factorized matrix elements are then expressed in terms of form
factors and decay constants. In this manner, the decay amplitudes can
be computed as a function of the parameters \cite{BAUER}
\begin{eqnarray}
  \label{eq:a1}
  a_1(\mu) \equiv C_1(\mu) + \frac{1}{N} C_2(\mu), \\
  a_2(\mu) \equiv C_2(\mu) + \frac{1}{N} C_1(\mu). \label{eq:a2} 
\end{eqnarray}
Analogous parameters $a_3$--$a_{10}$ can be introduced for the
contributions of QCD penguin operators $Q_{3-6}$ and electroweak
penguin operators $Q_{7-10}$ (see ref.~\cite{NLOQCD} for the
definition of the basis of operators). Since $C_2$ is of $O(1/N)$ with
respect to $C_1$, and has opposite sign, the two terms on the r.h.s.
of eq.~(\ref{eq:a2}) tend to cancel each other, and therefore one
finds that $a_2 \ll a_1$. Channels governed by $a_1$ are usually
called ``colour allowed'', while transitions governed by $a_2$ are
called ``colour suppressed''. Care must however be taken when trying
to quantify the effectiveness of colour suppression. Indeed, it
strongly depends on the actual value of the Wilson coefficients and on
the relative phase between the matrix elements of $Q_1$ and $Q_2$.

Factorization has been extensively used in phenomenological analyses
of $B$ decays \cite{BAUER}--\cite{LNF}, and it has proven
successful in estimating tree-dominated nonleptonic $B$ decays.
However, from the theoretical point of view, it is clear that
factorization cannot be exact. First of all, the factorized matrix
element, being expressed in terms of form factors and decay constants,
is scale- and scheme-independent, while the Wilson coefficients do
depend on the renormalization scale and scheme. Therefore,
eq.~(\ref{eq:bdp}) shows that any decay amplitude computed with the
factorization approximation carries these unphysical dependencies.
Furthermore, the neglect of FSI phases and nonfactorizable
contributions cannot in general be justified.  Finally, it should also
be stressed that the predictions within the factorization approach
suffer from a considerable model dependence in the computation of the
relevant form factors \cite{rom}.

\subsection{Generalized factorization}
\label{sec:genfac}

To overcome the problem of scale and scheme dependencies, two
generalizations of the factorization approach have been proposed. 

In the formulation of Neubert and Stech \cite{NS97}, the full matrix
elements are split into the factorized expression plus a
nonperturbative parameter $\varepsilon$: 
\begin{eqnarray}
&&\varepsilon^{(BD,\pi)}_1(\mu) \equiv  \frac{\langle \pi^- D^+ \vert
  (\bar d u)_{V-A} (\bar c b)_{V-A} \vert
  \bar B^0 \rangle}{\langle \pi^- D^+ \vert Q_1 \vert
  \bar B^0 \rangle_\mathrm{F}}-1,\nonumber \\
  &&\varepsilon^{(BD,\pi)}_8(\mu) \equiv  \frac{\langle \pi^- D^+
  \vert (\bar d t^a u)_{V-A} (\bar c t^a b)_{V-A}  \vert
  \bar B^0 \rangle}{2\langle \pi^- D^+ \vert Q_1 \vert
  \bar B^0 \rangle_\mathrm{F}}.\nonumber
\end{eqnarray}
$\varepsilon_{1,8}(\mu)$ combine with $C_{1,2}(\mu)$ to give the scale
and scheme independent parameters $a_1^\mathrm{eff}$ and
$a_2^\mathrm{eff}$. In this framework there is no explicit calculation of
non-factorizable contributions and $a_{1,2}^\mathrm{eff}$ are treated as
free parameters to be extracted from the data.

In refs.~\cite{GNF}--\cite{Soares}, it has been proposed to
improve on factorization by computing the $O(\alpha_s)$ corrections to
the matrix elements in perturbation theory:
\begin{equation}
  \label{eq:gi}
  \langle Q_i(\mu) \rangle \to g_i(\mu) \langle Q_i \rangle_\mathrm{T},
\end{equation}
where $\langle Q_i \rangle_\mathrm{T}$ denotes the tree-level
matrix element and $g_i(\mu)$ is a scheme and scale dependent
function. Then factorization is applied to $\langle Q_i
\rangle_\mathrm{T}$: 
\begin{eqnarray}
C_i(\mu) \langle Q_i(\mu) \rangle &\to& C_i(\mu) g_i(\mu) \langle Q_i
\rangle_\mathrm{T} \nonumber \\
&\to& C_i^\mathrm{eff} \langle Q_i
\rangle_\mathrm{F}.
  \label{eq:cieff}
\end{eqnarray}
The scheme and scale dependence of $g_i(\mu)$ matches that of
$C_i(\mu)$, so that the effective coefficients $C_i^\mathrm{eff}$ are
scheme and scale independent. Unfortunately, this result is obtained
at the price of losing control over finite terms in the effective
coefficients. Indeed, the $C_i^\mathrm{eff}$ computed with this recipe
depend on the choice of the gauge and of the external quark momenta
used in the perturbative evaluation of $g_i(\mu)$ \cite{BSI}.
To obtain effective coefficients that do not carry these 
dependencies and to give a physical meaning to $g_i(\mu)$, a
factorization theorem is necessary.

\subsection{Factorization theorems}
\label{sec:theorems}

The first step towards a factorization theorem was given by Bjorken's
colour transparency argument \cite{Bjorken}. Let us consider a decay
of the $B$ meson in two light pseudoscalars, where two light quarks
are emitted from the weak interaction vertex as a fast-traveling
small-size colour-singlet object. In the heavy-quark limit, soft
gluons cannot resolve this colour dipole and therefore soft gluon
exchange between the two light mesons decouples at lowest order in
$\Lambda_\mathrm{QCD}/m_b$.

Dugan and Grinstein \cite{DG} put forward a proof of factorization for
$B$ decays into a heavy-light final state, in which the light meson is
emitted. Unfortunately, their proof is based on the use of the
so-called ``large energy effective theory'', which is known to be
unsuitable to describe exclusive processes, since it fails to
consistently account for the hadronization of the emitted meson
\cite{Aglietti:1992ta}. The large energy effective theory can only be
used to prove factorization for semi-inclusive processes \cite{ac}.

Politzer and Wise \cite{PW} have applied the Brodsky and Lepage
formalism \cite{BL} to write a factorization formula for nonleptonic
$b \to c$ transitions in which a light meson is emitted and the
spectator quark is absorbed by the charmed meson. The expression,
valid in the limit of heavy $b$ and $c$ quarks with $r=m_c/m_b$ fixed,
reads:
\begin{eqnarray}
  \label{eq:wise}
  && \langle H_c(v^\prime) \pi(P)  \vert Q_i(m_b) \vert H_b(v) \rangle = \\
  && \frac{1}{4}\langle H_c(v^\prime)  \vert \bar h_{v^\prime}^{(c)}
  h_{v}^{(b)} \vert H_b(v) \rangle m_b f_\pi (1-r) \times \nonumber \\
  &&\int_0^1 \mathrm{d}x \,
  T_i^\mathrm{(S)} (x,r,m_b) \phi_\pi(x,m_b) +\nonumber \\
  && \frac{1}{4}\langle H_c(v^\prime)  \vert \bar h_{v^\prime}^{(c)}
  \gamma_5 h_{v}^{(b)} \vert H_b(v) \rangle m_b f_\pi (1+r) \times
  \nonumber \\
  &&\int_0^1 \mathrm{d}x \,
  T_i^\mathrm{(P)} (x,r,m_b) \phi_\pi(x,m_b),\nonumber
\end{eqnarray}
where the renormalization scale has been set to $m_b$ and
$\phi_\pi(x,m_b)$ is the pion light-cone distribution amplitude. The
hard scattering amplitudes $T_i^\mathrm{(S,P)}(x,r,m_b)$ can be
perturbatively expanded in $\alpha_s(m_b)$ and are obtained from the
computation of the diagrams in fig.~\ref{fig:wise}. 

\FIGURE[t]{\epsfig{file=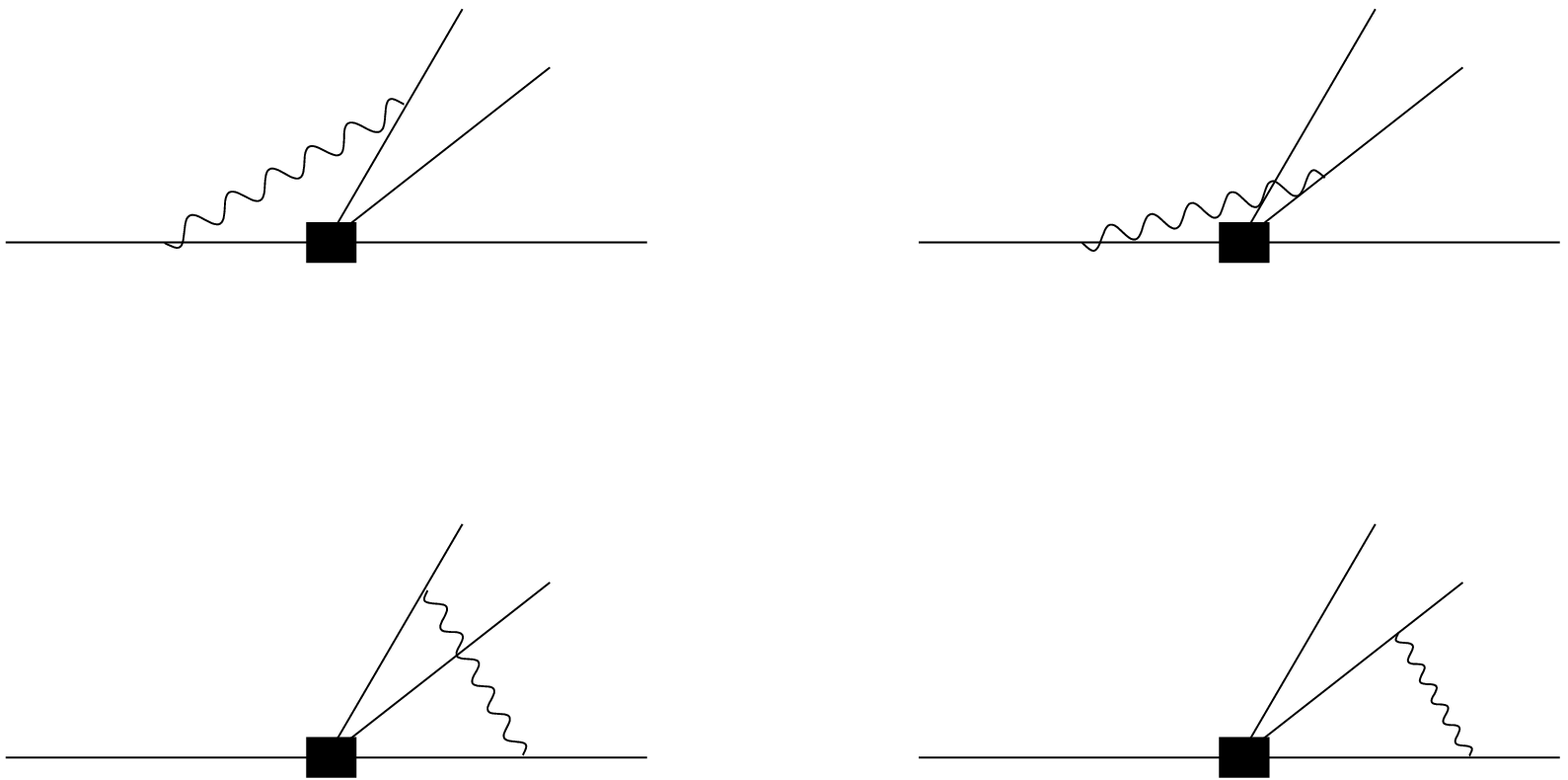,width=7cm}\caption{Feynman diagrams
    for the computation of the hard scattering amplitudes
    $T$.}\label{fig:wise}}

Beneke, Buchalla, Neubert and Sachrajda \cite{BBNS} have recently
extended this formulation to $B$ decays into two light mesons and
supplemented it with an explicit one-loop proof of factorization for
$B \to \pi \pi$ decays valid in the limit of a heavy $B$ meson.
Assuming that in $B \to \pi \pi$ decays perturbative Sudakov
suppression is not sufficient to guarantee the dominance of hard
spectator interactions, they argue that all soft spectator
interactions can be absorbed in the $B\to \pi$ form factor. They
therefore obtain the following factorization formula, valid at lowest
order in $\Lambda_\mathrm{QCD}/m_b$:
\begin{eqnarray}
  \label{eq:bbns}
 && \langle \pi(p^\prime) \pi(q) \vert Q_i\vert \bar B(p)\rangle = \\
 && f^{B \to \pi} (q^2) \int_0^1 \mathrm{d}x \, T_i^I(x)\phi_\pi(x) +
 \nonumber \\
 && \int_0^1 \mathrm{d}\xi \, \mathrm{d}x\, \mathrm{d}y\,
 T^{II}_i(\xi,x,y) \phi_B(\xi) \phi_\pi(x) \phi_\pi(y), \nonumber
\end{eqnarray}
where $f^{B \to \pi} (q^2)$ is a $B\to \pi$ form factor, and
$\phi_\pi$ $(\phi_B)$ are leading-twist light cone distribution
amplitudes of the pion ($B$ meson). Analogously to
eq.~(\ref{eq:wise}),  $T^{I,II}_i$ denote the hard scattering
amplitudes. It is important to notice that $T^I$ starts at zeroth
order in $\alpha_s$, giving factorization as in eq.~(\ref{eq:fact1}),
and at higher order contains hard gluon exchange not involving the
spectator, while $T^{II}$ contains the hard interactions of the
spectator and starts at order $\alpha_s$ (see fig.~\ref{fig:beneke}).
This implies that factorization formulae for $B\to \pi\pi$ based on
the dominance of hard spectator interactions \cite{SHB,taiwan} miss
the leading contribution in $\alpha_s$, unless Sudakov suppression is
so effective that hard interactions of the spectator are dominant over
soft ones.  This controversial point should be clarified in order to
determine the relative weight of the two terms on the r.h.s. in
eq.~(\ref{eq:bbns}).

\FIGURE[t]{\epsfig{file=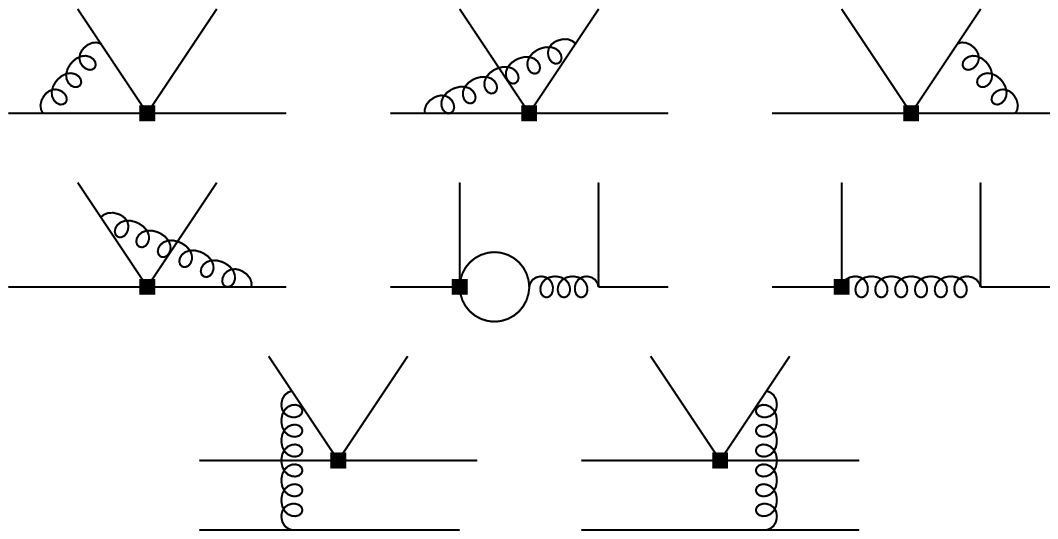,width=7cm}\caption{Order
    $\alpha_s$ corrections to the hard scattering kernels $T_i^I$
    (first two rows) and $T_i^{II}$ (third row).}\label{fig:beneke}}

In the case where the spectator is absorbed by a heavy meson, the
second term on the r.h.s.  in eq.~(\ref{eq:bbns}) is power suppressed
(i.e. hard interactions of the spectator are power suppressed)
and one recovers eq.~(\ref{eq:wise}). Finally, if a heavy
meson is emitted, factorization cannot be justified.\footnote{These
  considerations cast serious doubts on the validity of the factorization
  formulae of ref.~\cite{taiwan} and on the phenomenological analysis
  of ref.~\cite{chengnew}.}

The scheme and scale dependence of the scattering kernels $T^{I,II}_i$
matches the one of Wilson coefficients, and the final result is
consistently scale and scheme independent. 

Final state interaction phases appear in this formalism as imaginary
parts of the scattering kernels (at lowest order in
$\Lambda_\mathrm{QCD}/m_b$). These phases appear in the computation of
penguin contractions and of hard gluon exchange between the two pions.
This means that in the heavy quark limit final state interactions can
be determined perturbatively.

The formalism of ref.~\cite{BBNS} is certainly a very interesting
theoretical result. From the point of view of phenomenological
applications, however, the issue of $\Lambda_\mathrm{QCD}/m_b$
corrections is still an open problem that deserves further
investigation, especially in those cases where
$\Lambda_\mathrm{QCD}/m_b$ corrections are chirally or Cabibbo enhanced.

\section{General Parameterizations}
\label{sec:genpar}

In order to identify possibly dangerous nonfactorizable contributions
and to estimate the uncertainties in the extraction of standard model
parameters from nonleptonic $B$ decays, it is useful to complement the
above approach developing a general parameterization of decay
amplitudes. 

\subsection{The ``diagrammatic'' approach}
\label{sec:diag}

The diagrammatic approach \cite{diagr1,diagr2} has been extensively
used to describe CP violating $B$ decays. It is based on the
flavour-flow topologies of Feynman diagrams in the full theory. 

In the approach of ref.~\cite{diagr2} the basic parameters for
strangeness preserving decays are the ``Tree'' (colour favoured)
amplitude $T$, the ``colour suppressed'' amplitude $C$, the
``penguin'' amplitude $P$, the ``exchange'' amplitude $E$, the
``annihilation'' amplitude $A$ and the ``penguin annihilation''
amplitude ${\it PA}$. For strangeness changing decays
one has $T^\prime$, $C^\prime$, $P^\prime$, $E^\prime$, $A^\prime$,
${\it PA}^\prime$. If $Z$-penguins are taken into account one
introduces in addition the ``colour-allowed'' $Z$-penguin $P_{\sss
  {\rm EW}}$ and the ``colour-suppressed'' $Z$-penguin $P^C_{\sss {\rm
    EW}}$. Similarly $P_{\sss {\rm EW}}^\prime$ and $P_{\sss {\rm
    EW}}^{\prime C}$ are introduced for strangeness-changing decays.

Recently the usefulness of the diagrammatic approach has been
questioned with respect to the effects of final state interactions. In
particular various ``plausible'' diagrammatic arguments to neglect
certain flavour-flow topologies may not hold in the presence of FSI,
which mix up different classes of diagrams
\cite{FSI}--\cite{CHARMING1}.  Another criticism which one may add is
the lack of an explicit relation of this approach to the basic
framework for non-leptonic decays represented by the effective weak
Hamiltonian and OPE in eq.~(\ref{eq:ope}). In particular, the
diagrammatic approach is governed by Feynman drawings with $W$-, $Z$-
and top-quark exchanges.  Yet such Feynman diagrams with full
propagators of heavy fields represent really the situation at very
short distance scales $O(M_{W,Z},m_t)$, whereas the true picture of a
decaying meson with a mass $O(m_b)$ is more properly described by
effective point-like vertices represented by the local operators
$Q_i$. The effect of $W,Z$ and top quark exchanges is then described
by the values of the Wilson coefficients of these operators. The only
explicit fundamental degrees of freedom in the effective theory are
the quarks $u$, $d$, $s$, $c$, $b$, the gluons and the photon.

In view of this situation it is desirable to develop another
phenomenological approach based directly on the OPE which allows a
systematic description of non-factorizable contributions such as
penguin contributions and final state interactions. Simultaneously one
would like to have an approach that does not lose the intuition of the
diagrammatic approach while avoiding the limitations of the latter.

\subsection{Wick contractions and charming penguins}
\label{sec:charming}

A general parameterization which fulfills the above requirements has
been proposed in ref.~\cite{CHARMING1}. This approach is based on
identifying the different topologies of Wick contractions in the
matrix elements of the operators $Q_i$ (see figs.~\ref{fig:emission}
and \ref{fig:penguin}). There are emission topologies: Disconnected
Emission ({\it DE}\/) and Connected Emission ({\it CE\/});
annihilation topologies: Disconnected Annihilation ({\it DA\/}) and
Connected Annihilation ({\it CA\/}); emission annihilation topologies:
Disconnected Emission Annihilation ({\it DEA}\/) and Connected
Emission Annihilation ({\it CEA\/}); penguin topologies: Disconnected
Penguin ({\it DP\/}) and Connected Penguin ({\it CP\/});
penguin emission topologies: Disconnected Penguin Emission ({\it
  DPE\/}) and Connected Penguin Emission ({\it CPE\/});
penguin annihilation topologies: Disconnected Penguin Annihilation
({\it DPA\/}) and Connected Penguin Annihilation ({\it CPA\/});
double penguin annihilation topologies: Disconnected
Double Penguin Annihilation ($\overline{\it DPA}$) and Connected
Double Penguin Annihilation ($\overline{\it CPA}$). The dashed lines
represent the operators.  All these parameters are flavour dependent,
and some symmetry argument or dynamical assumption is needed in
general to relate parameters entering different decay channels. The
apparently disjoint pieces in the topologies {\it DEA}, {\it CEA},
{\it DPE}, {\it CPE}, {\it DPA}, {\it CPA}, $\overline{\it DPA}$ and
$\overline{\it CPA}$ are connected to each other by gluons or photons,
which are not explicitly shown. These special topologies in which only
gluons connect the disjoint pieces are Zweig suppressed and are
therefore naively expected to play a minor role in $B$ decays.
However, they have to be included in order to define scheme and scale
independent combinations of Wilson coefficients and matrix elements
(see Section \ref{sec:BS}).

\FIGURE[t]{\epsfig{file=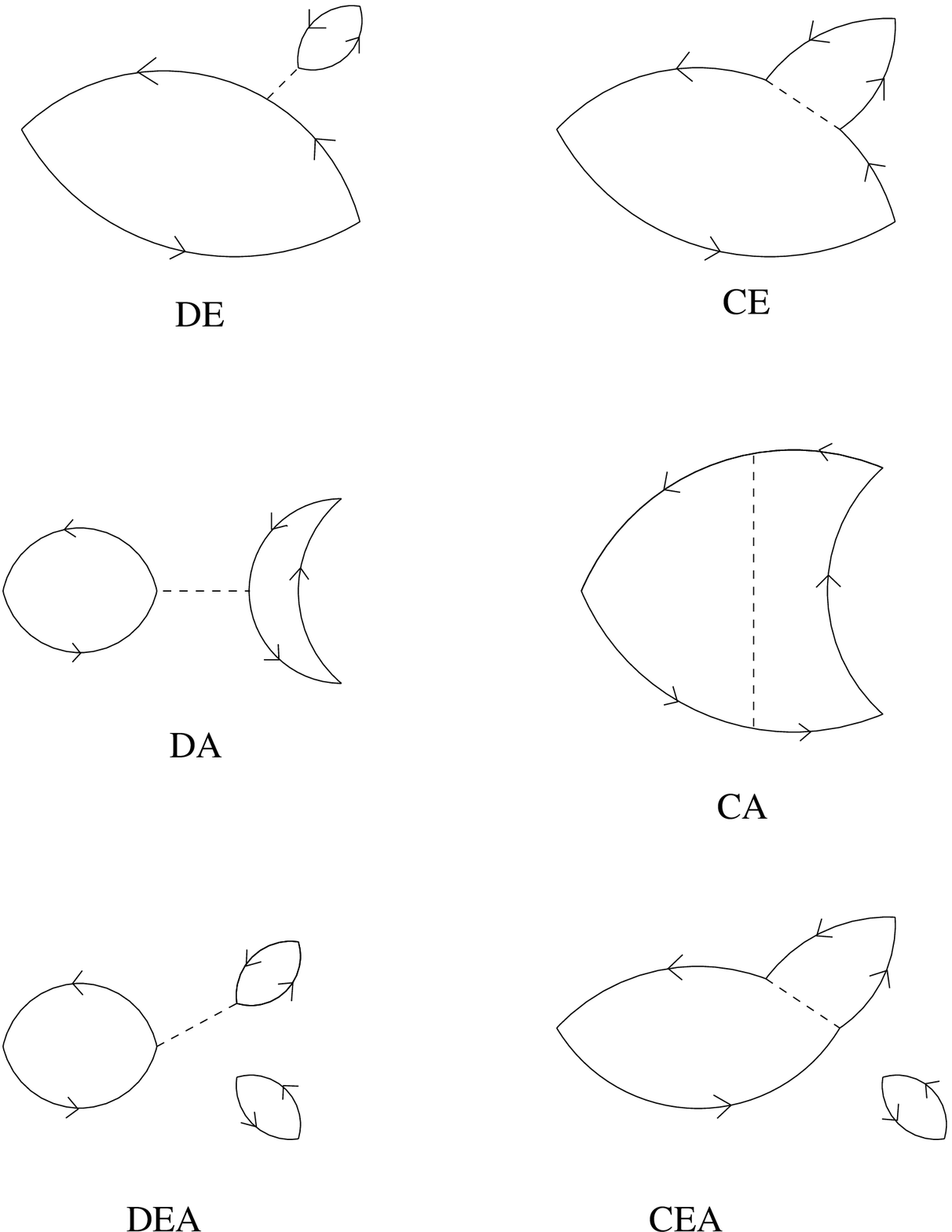,width=7cm}\caption{Emission and
    annihilation topologies of Wick contractions in the matrix
    elements of operators $Q_i$.}\label{fig:emission}}
\FIGURE[t]{\epsfig{file=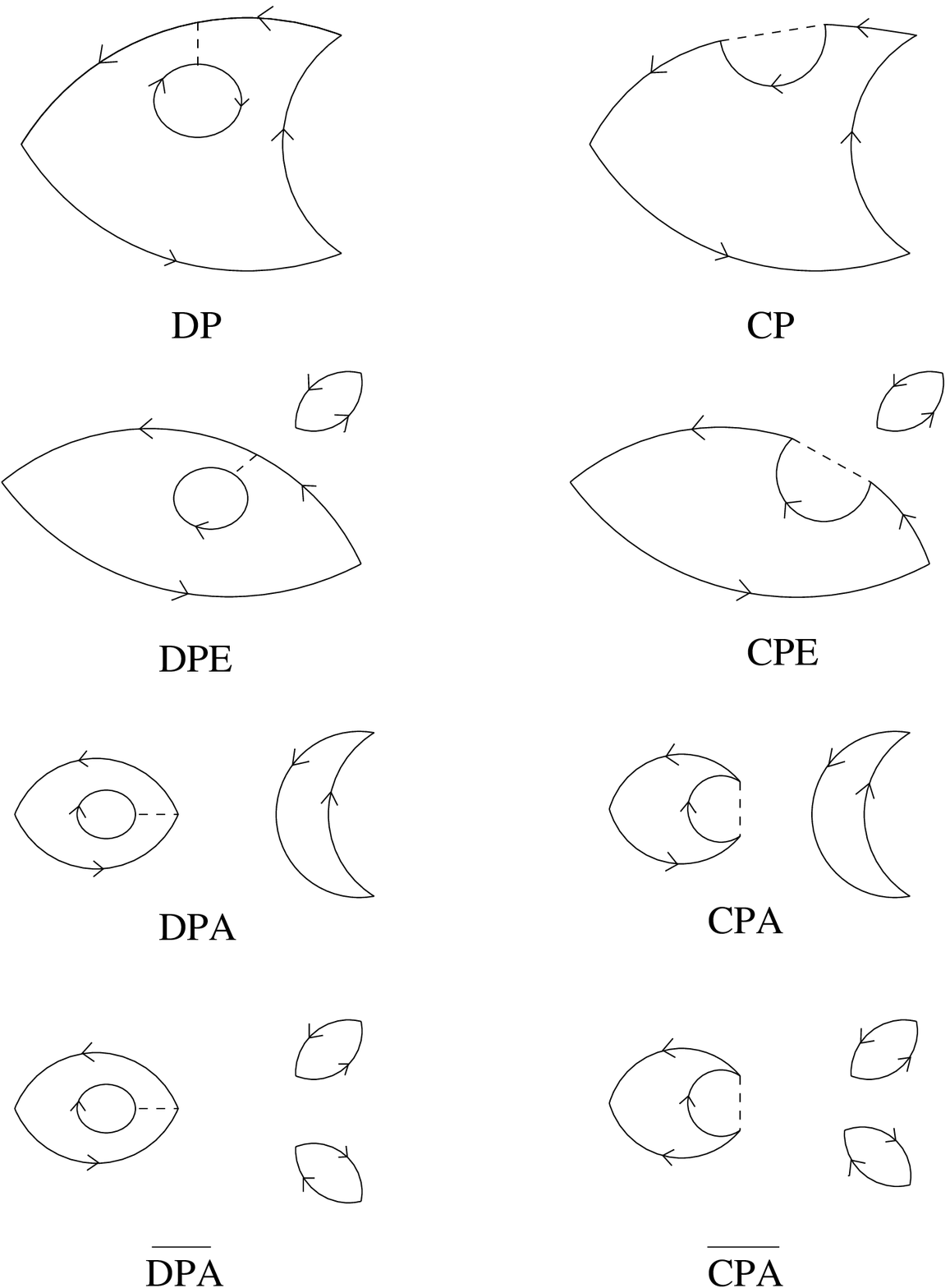,width=7cm}\caption{Penguin topologies
    of Wick contractions in the matrix elements of operators
    $Q_i$.}\label{fig:penguin}}

The Wick contractions are complex because of final state
interactions. For example, it is easy to show that a disconnected
emission followed by a rescattering is equivalent to a connected
penguin. The same holds for annihilations. Therefore, by computing all
possible Wick contractions one is automatically taking into account
all rescattering effects in a consistent way.

The decay amplitude can be readily identified as a sum of Wilson
coefficients times Wick contractions.  These Wick contractions might
eventually be computed on the lattice, with the caveat discussed in
the Introduction.  Letting the parameters vary in reasonable ranges,
it is possible to obtain a reliable estimate of the uncertainties in
the extraction of CKM parameters from CP asymmetries in $B$ decays.

It is interesting to note that some particular topologies, the
so-called ``charming penguins'', i.e.  penguin contractions with a
$c$-quark loop, can give very large contributions to some decay
channels, even if their absolute value is very small. A typical
example is given by $B \to K \pi$ decays \cite{CHARMING1,CHARMING2},
where the following contributions are present (neglecting
annihilations and GIM-suppressed penguins):
\begin{enumerate}
\item The contribution of emission matrix elements of current-current
  operators containing up quarks, which is doubly Cabibbo suppressed,
  but has Wilson coefficients and matrix elements of $O(1)$ (we
  normalize all matrix elements to \textit{DE} for simplicity):
  \begin{eqnarray}
   && V_{us} V_{ub}^*  \left(C_1 \langle Q_1^u\rangle_\mathit{DE} + C_2
    \langle Q_2^u \rangle_\mathit{CE} \right)\nonumber \\
  &&\qquad \sim O(\lambda^4 \cdot 1
    \cdot 1);
    \label{eq:cont1}
  \end{eqnarray}

\item The contribution of emission matrix elements of penguin
  operators, which is Cabibbo allowed, but has Wilson coefficients of
  $O(10^{-2})$ and matrix elements of $O(1)$:
  \begin{eqnarray}
    && V_{ts} V_{tb}^* \sum_{i=2}^5 \left(C_{2i-1} \langle
    Q_{2i-1}\rangle_\mathit{DE} + C_{2i} 
    \langle Q_{2i} \rangle_\mathit{CE} \right) \nonumber \\
  &&\qquad \sim O(\lambda^2 \cdot 10^{-2}
    \cdot 1);
    \label{eq:cont2}
  \end{eqnarray}
  
\item The contribution of penguin contractions of current-current
  operators involving the charm quark (charming penguins), which is
  Cabibbo allowed, has Wilson coefficients of $O(1)$ and small
  ($O(\Lambda_\mathrm{QCD}/m_b)$ in the framework of ref.~\cite{BBNS})
  matrix elements:
  \begin{eqnarray}
   && V_{ts} V_{tb}^* \left(C_1 \langle Q_1^c\rangle_\mathit{DP} + C_2
    \langle Q_2^c \rangle_\mathit{CP} \right) \nonumber \\
  &&\qquad \sim O(\lambda^2 \cdot 1
    \cdot \frac{\Lambda_\mathrm{QCD}}{m_b});
    \label{eq:cont3}
  \end{eqnarray}

\item The contribution of penguin contractions of penguin operators,
  which is Cabibbo allowed but has small Wilson coefficients and
  matrix elements:
  \begin{eqnarray}
    && V_{ts} V_{tb}^* \sum_{i=2}^5 \left(C_{2i-1} \langle
    Q_{2i-1}\rangle_\mathit{DP} + C_{2i} 
    \langle Q_{2i} \rangle_\mathit{CP} \right) \nonumber \\
  &&\qquad \sim O(\lambda^2 \cdot 10^{-2}
    \cdot \frac{\Lambda_\mathrm{QCD}}{m_b}).
    \label{eq:cont4}
  \end{eqnarray}
\end{enumerate}

It is evident that the contribution of charming penguins is likely to
be large even for values of the penguin contractions of
$O(\Lambda_\mathrm{QCD}/m_b)$, thanks to the Cabibbo enhancement and
the large Wilson coefficient. This means that great care should be
taken in applying the factorization approach of ref.~\cite{BBNS} to
these channels, and particularly in estimating hadronic uncertainties
in the extraction of the angle $\gamma$ from these decays
\cite{CLEOg,gamman}.

\subsection{A RGE invariant parameterization}
\label{sec:BS}

One disadvantage of the above formalism is that the Wick contraction
parameters $DE$, $CE$ etc., individually depend on the renormalization
scale and scheme and they combine in a rather complicated way with
the Wilson coefficients to give a scheme and scale independent
result. It is however possible to identify a complete set of scale and
scheme independent parameters, made up of combinations of Wilson
coefficients and Wick parameters \cite{BSII}. This means that it is
possible to give an expression for the decay amplitude that is
manifestly RGE invariant and retains all the advantages of the
effective Hamiltonian approach. There is a correspondence
of these parameters with the parameters of the ``diagrammatic''
approach: in some sense, these RGE invariant parameters are the
translation in the effective Hamiltonian language of the parameters of
the diagrammatic approach. 

I now briefly describe the RGE invariant effective parameters. The
interested reader will find the complete definition of the effective
parameters in ref.~\cite{BSII}, together with a detailed derivation.

The first scale and scheme independent combinations of Wilson
coefficients and matrix elements that one can identify correspond to
the $T$ and $C$ parameters of the diagrammatic approach. Denoting by
$\langle Q_i \rangle_{DE}$ and $\langle Q_i \rangle_{CE}$ the
insertions of $Q_i$ into ${\it DE}$ and ${\it CE}$ topologies
respectively, one finds two effective parameters
\begin{eqnarray}
  E_1&=& C_1\, \langle Q_1 \rangle_{{\it DE}} + C_2 \,\langle Q_2 
  \rangle_{{\it CE}}
  \,, \nonumber \\
  E_2&=& C_1\, \langle Q_1 \rangle_{{\it CE}} + C_2 \,\langle Q_2 
  \rangle_{{\it DE}}\,.
  \label{eq:E1E20}
\end{eqnarray}
For simplicity, we suppress here and in the following the flavour
variables. They are given explicitly in ref.~\cite{BSII}. $E_1$ and
$E_2$ are generalizations of $a_1^{\rm eff}\langle Q_1 \rangle_F$ and
$a_2^{\rm eff}\langle Q_2 \rangle_F$ in the formulation of
ref.~\cite{NS97}.

Next, annihilation topologies must be considered. The $A$ and $E$
parameters of the diagrammatic approach correspond respectively to the
effective parameters
\begin{eqnarray}
  A_1&=& C_1\, \langle Q_1 \rangle_{\it DA} + C_2 \,\langle Q_2 
  \rangle_{\it CA}
  \,, \nonumber \\
  A_2&=& C_1\, \langle Q_1 \rangle_{\it CA} + C_2 \,\langle Q_2 
  \rangle_{\it DA}\,,
  \label{eq:A1A20}
\end{eqnarray}
where $\langle Q_i \rangle_{\it DA}$ and $\langle Q_i \rangle_{\it
  CA}$ denote the $Q_i$-insertions into ${\it DA}$ and ${\it CA}$
topologies respectively. Due to the flavour structure of operators
$Q_1$ and $Q_2$, $A_1$ can only contribute to $B^+$ decays while $A_2$
can only contribute to $B_{d,s}$ decays.

The last class of non-penguin contractions that we consider
corresponds to the insertion of $Q_1$ and $Q_2$ into
emission-annihilation topologies, denoted by ${\it DEA}$ and ${\it
  CEA}$ in fig.~\ref{fig:emission}. Proceeding as above, we can
identify two new effective parameters:
\begin{eqnarray}
  {\it EA}_1&=& C_1\, \langle Q_1 \rangle_{\it DEA} + C_2 \,\langle Q_2 
  \rangle_{\it CEA}, \nonumber \\
  {\it EA}_2&=& C_1\, \langle Q_1 \rangle_{\it CEA} + C_2 \,\langle Q_2 
  \rangle_{\it DEA}.
  \label{eq:EA1EA20}
\end{eqnarray}
As in the case of $A_1$ and $A_2$, due to the flavour structure of
$Q_1$ and $Q_2$, ${\it EA}_1$ can only contribute to $B^+$ decays while
${\it EA}_2$ can only contribute to $B_{d,s}$ decays.

The situation with penguin contractions is a little bit more involved.
Indeed, to obtain scale and scheme independent effective parameters it
is necessary to combine penguin contractions of current-current
operators and matrix elements of QCD and electroweak penguin operators.

Similarly to the sets $(E_1,E_2)$, $(A_1,A_2)$ and $({\it
  EA}_1,{\it EA}_2)$ one can find four effective
``penguin''-parameters $P_1$, $P_2$, $P_3$ and $P_4$:
\begin{enumerate}
\item $P_1$ involves the insertions of $Q_1$ and $Q_2$ into ${\it CP}$
  and ${\it DP}$ topologies respectively and a particular set of
  matrix elements of QCD-penguin and electroweak penguin operators
  necessary for the cancellation of scale and scheme dependences;
  
\item $P_2$ involves the insertions of $Q_1$ and $Q_2$ into ${\it
    CPE}$ and ${\it DPE}$ topologies respectively and a suitable set
  of matrix elements of QCD-penguin and electroweak penguin operators
  necessary for the cancellation of scale and scheme dependences;
  
\item $P_3$ involves the insertions of $Q_1$ and $Q_2$ into ${\it
    CPA}$ and ${\it DPA}$ topologies respectively and the
  corresponding set of matrix elements of QCD-penguin and electroweak
  penguin operators necessary for the cancellation of scale and scheme
  dependences;
  
\item $P_4$ involves the insertions of $Q_1$ and $Q_2$ into
  $\overline{\it CPA}$ and $\overline{\it DPA}$ topologies
  respectively and the remaining matrix elements of QCD-penguin and
  electroweak penguin operators which have not been included in $P_1$,
  $P_2$ and $P_3$.
\end{enumerate}

The explicit expressions for the $P_1$, $P_2$, $P_3$ and $P_4$
parameters are as follows:
\begin{eqnarray}
  P_1&=&C_1 \langle Q_1\rangle^c_{\it CP} + 
  C_2 \langle Q_2\rangle^c_{DP} \nonumber \\ 
  &+&\sum_{i=2}^5 \Bigl(C_{2i-1} \langle Q_{2i-1} \rangle_{{\it CE}}
  + C_{2i} \langle Q_{2i} \rangle_{{\it DE}}\Bigr) \nonumber \\
  &+&\sum_{i=3}^{10} \Bigl(C_{i} \langle Q_{i} \rangle_{\it CP}
  + C_{i} \langle Q_{i} \rangle_{DP}\Bigr)\nonumber \\   
  &+&\sum_{i=2}^5 \Bigl(C_{2i-1} \langle Q_{2i-1} \rangle_{\it CA}
  + C_{2i} \langle Q_{2i} \rangle_{\it DA}\Bigr)\, , \nonumber \\
  P_2&=&C_1 \langle Q_1 \rangle^c_{\it CPE} + 
  C_2 \langle Q_2 \rangle^c_{\it DPE} \nonumber \\ 
  &+& \sum_{i=2}^5 \Bigl(C_{2i-1} \langle Q_{2i-1} \rangle_{{\it DE}}
  + C_{2i} \langle Q_{2i} \rangle_{{\it CE}}\Bigr) \nonumber \\
  &+&\sum_{i=2}^5 \Bigl(C_{2i-1} \langle Q_{2i-1} \rangle_{{\it CEA}}
  + C_{2i} \langle Q_{2i} \rangle_{{\it DEA}}\Bigr)\nonumber \\ 
  &+& \sum_{i=3}^{10} \Bigl(C_{i} \langle Q_{i} \rangle_{\it CPE}
  + C_{i} \langle Q_{i} \rangle_{\it DPE}\Bigr)\, , \nonumber \\
  P_3&=&C_1 \langle Q_1 \rangle^c_{\it CPA} + 
  C_2 \langle Q_2 \rangle^c_{\it DPA} \nonumber \\ 
  &+& \sum_{i=2}^5 \Bigl(C_{2i-1} \langle Q_{2i-1} \rangle_{\it DA}
  + C_{2i} \langle Q_{2i} \rangle_{\it CA}\Bigr) \nonumber \\
  &+&\sum_{i=3}^{10} \Bigl(C_{i} \langle Q_{i} \rangle_{\it CPA}
  + C_{i} \langle Q_{i} \rangle_{\it DPA}\Bigr)\, , \nonumber \\
  P_4&=&C_1 \langle Q_1 \rangle^c_{\overline{\it CPA}} + 
  C_2 \langle Q_2 \rangle^c_{\overline{\it DPA}} \nonumber \\ 
  &+& \sum_{i=2}^5 \Bigl(C_{2i-1} \langle Q_{2i-1} \rangle_{\it DEA}
  + C_{2i} \langle Q_{2i} \rangle_{\it CEA}\Bigr) \nonumber \\
  &+&\sum_{i=3}^{10} \Bigl(C_{i} \langle Q_{i} \rangle_{\overline{\it CPA}}
  + C_{i} \langle Q_{i} \rangle_{\overline{\it DPA}}\Bigr)\, ,
  \label{eq:p4gen}
\end{eqnarray}
where we have denoted by $\langle Q_i\rangle^c_{\it CP}$ the insertion
of operator $Q_i$ in a {\it CP\/} topology with a $c$-quark running in
the loop (this corresponds to the charming penguin of
ref.~\cite{CHARMING1}), and analogously for ${\it DP}$, ${\it CPE}$,
${\it DPE}$, ${\it CPA}$, ${\it DPA}$, $\overline{\it CPA}$ and
$\overline{\it DPA}$ topologies. Notice that, due to the flavour
structure of the penguin-annihilation contributions, $P_3$ and $P_4$
cannot contribute to $B^+$ decays. Moreover $P_4$ contributes only to
final states with two flavour neutral mesons $\bar q_1 q_1$ and $\bar
q_2 q_2$. Similarly $P_2$ contributes only to states with at least one
flavour neutral meson $\bar q_2 q_2$. 

The $P_i$ parameters are always accompanied by
the CKM factor $V_{tb} V_{td_i}^*$, where $d_i={d,s}$. Penguin-type
matrix elements are also present in the part of ${\cal H}_{\rm eff}$
proportional to $V_{ub} V_{ud_i}^*$. They correspond to penguin
contractions of operators $Q_{1,2}$, or, more precisely, of the
differences $\Bigl(Q_1^{d_iuu}-Q_1^{d_icc}\Bigr)$ and
$\Bigl(Q_2^{d_iuu}-Q_2^{d_icc}\Bigr)$. When these combinations are
inserted into penguin topologies, they give rise to a generalization
of the GIM penguins of ref.~\cite{CHARMING1}. The scale and scheme
independent contributions are given by
\begin{eqnarray}
  P_1^{\sss {\rm GIM}}&=& C_1 \Bigl(\langle Q_1 \rangle^c_{\it CP} -
  \langle Q_1 
  \rangle^u_{\it CP} \Bigr) \nonumber \\ 
  &+& C_2 \Bigl(\langle Q_2 \rangle^c_{DP} - \langle Q_2
  \rangle^u_{DP} \Bigr)\,,\nonumber \\
  P_2^{\sss {\rm GIM}}&=& C_1 \Bigl(\langle Q_1 \rangle^c_{\it CPE} -
  \langle Q_1 
  \rangle^u_{\it CPE} \Bigr) \nonumber \\ 
  &+& C_2 \Bigl(\langle Q_2 \rangle^c_{\it DPE} - \langle Q_2
  \rangle^u_{\it DPE} \Bigr)\,,\nonumber \\
  P_3^{\sss {\rm GIM}}&=& C_1 \Bigl(\langle Q_1 \rangle^c_{\it CPA} -
  \langle Q_1 
  \rangle^u_{\it CPA} \Bigr) \nonumber \\ 
  &+& C_2 \Bigl(\langle Q_2 \rangle^c_{\it DPA} - \langle Q_2
  \rangle^u_{\it DPA} \Bigr)\,,\nonumber \\
  P_4^{\sss {\rm GIM}}&=& C_1 \Bigl(\langle Q_1
  \rangle^c_{\overline{\it CPA}} - 
  \langle Q_1 
  \rangle^u_{\overline{\it CPA}} \Bigr) \nonumber \\ 
  &+& C_2 \Bigl(\langle Q_2 \rangle^c_{\overline{\it DPA}} - \langle Q_2
  \rangle^u_{\overline{\it DPA}} \Bigr)\,,
  \label{eq:gengim}
\end{eqnarray}
and vanish in the limit of degenerate $u$ and $c$.  The unitarity
of the CKM matrix assures that in a given decay $P_i$ is always
accompanied by $P_i^{\sss {\rm GIM}}$. However, $P_i$ and $P_i^{\sss
  {\rm GIM}}$ are always multiplied by different CKM factors and in
order to study CP violation it is more convenient to keep the latter
factors explicitly and consider separately $P_i$ and $P_i^{\sss {\rm
    GIM}}$.

The relation to the parameters of the diagrammatic approach is the
following:
\begin{eqnarray}
  P_c - P_t + P^c_{\sss {\rm EW}} - P^t_{\sss {\rm EW}} 
  &\leftrightarrow& P_1,\nonumber \\ 
  P_c - P_u + P^c_{\sss {\rm EW}} - P^u_{\sss {\rm EW}} 
  &\leftrightarrow& P_1^{\sss
  {\rm GIM}}\,.
  \label{eq:pcptew}
\end{eqnarray}

A hierarchy between the effective parameters can be established with
the help of some dynamical considerations. As an example, I report
here the results of a large $N$ classification \cite{BSII}. One has in
units of $\sqrt{N}$ the following hierarchy for various topologies:
\begin{eqnarray}
  \label{eq:hier1}
  {\it DE},\,{\it DA}&:& \,\,O (1), \\
  {\it CE},\,{\it CA},\,{\it DEA},\,{\it CP},\,{\it DPA},\,{\it
  DPE}&:&\,\, O(1/N), \nonumber \\
  {\it CEA},\,{\it DP},\,{\it CPA},\,{\it CPE},\,{\it
  \overline{DPA}}&:&\,\, O(1/N^2), \nonumber \\ 
  {\it \overline{CPA}}&:&\,\, O (1/N^3). \nonumber
\end{eqnarray}
Combining these results with the large $N$ hierarchy of Wilson
coefficients one gets the following 
classification of the effective parameters in units of $\sqrt{N}$:
\begin{eqnarray}
  \label{eq:classhier}
  E_1,\,\, A_1&:& \,\, O(1),\\
  E_2,\,\,A_2,\,\,{\it EA}_1,\,\,P_1,\,\,P_1^{\sss {\rm
  GIM}}&:&\,\,O(1/N), \nonumber\\
  {\it EA}_2,\,\,P_2,\,\,P_3,\,\,P_2^{\sss {\rm
  GIM}},\,\,P_3^{\sss {\rm
  GIM}}&:&\,\,O(1/N^2), \nonumber\\
  P_4,\,\,P_4^{\sss {\rm
  GIM}}&:&\,\, O(1/N^3). \nonumber
\end{eqnarray}

Using the above considerations, it is possible to divide two-body
nonleptonic $B$ decay channels in various classes, according to the
CKM structure and to the effective parameters entering the decay. This
classification has been given in ref.~\cite{BSII}, together with the
full expressions of the decay amplitudes in terms of the effective
parameters. 

\section{Conclusions and outlook}
\label{sec:concl}

As I have summarized in the previous sections, some interesting
progress has been recently made in our understanding of nonleptonic
$B$ decays. Factorization in the limit of a heavy $B$ meson has been
proven at one loop for $B$ decays to two light pseudoscalars
\cite{BBNS}, and it can be extended to decays to a heavy-light state
where the spectator is absorbed by the heavy meson. At the lowest
order in $\Lambda_\mathrm{QCD}/m_b$ corrections to the factorization
approximation can be systematically computed in perturbation
theory. In the same approximation, FSI can be computed perturbatively.
These theoretically very exciting results should be confronted with
phenomenology keeping in mind that nonfactorizable power corrections
can in some cases be chirally or Cabibbo enhanced. 

To this aim, general parameterizations of decay amplitudes can be very
helpful in identifying the potentially dangerous contributions. For
example, the Wick contraction parameterization of
ref.~\cite{CHARMING1}, tightly connected to the OPE formalism, shows
that a particular class of contributions, the ``charming penguins''
(penguin contractions of charmed current-current operators), can give
large, if not dominant, contribution to many interesting decay
channels such as $B \to K \pi$.

In its RGE invariant formulation \cite{BSII}, the Wick contraction
parameterization benefits of the advantages of being directly
connected to the effective Hamiltonian formalism, while maintaining
the appealing simplicity of the diagrammatic approach.

It is very exciting to think that benefiting of these complementary
approaches as well as of the wealth of forthcoming data we might be
able in the near future to reach a good understanding of nonleptonic
$B$ decays.

\acknowledgments

I warmly thank the organizers of this very interesting and
stimulating conference.  I am much indebted to A.J. Buras, M.
Ciuchini, R. Contino, E. Franco and G. Martinelli for the lively and
enjoyable collaboration on many of the subjects discussed in this talk.
I acknowledge the support of the German Bundesministerium f{\"u}r
Bildung und Forschung under contract 05 HT9WOA.

\end{document}